# Effects of the sintering atmosphere on the superconductivity of SmFeAsO$_{1-x}$F$_x$ compounds

Y. Ding, Y. Sun, X. D. Wang, J. C. Zhuang, L. J. Chui, Z. X. Shi*
Corresponding author Z. X. Shi: zxshi@seu.edu.cn
Department of Physics, Southeast University, Nanjing 211189, P. R. China

**Abstract**

A series of SmFeAsO$_{1-x}$F$_x$ samples were sintered in quartz tubes filled with air of different pressures. The effects of the sintering atmosphere on the superconductivity were systematically investigated. The SmFeAsO$_{1-x}$F$_x$ system maintains a transition temperature ($T_c$) near 50 K until the concentration of oxygen in quartz tubes increases to a certain threshold, after which $T_c$ decreases dramatically. Fluorine losses, whether due to vaporization, reactions with starting materials, and reactions with oxygen, proved to be detrimental to the superconductivity of this material. The deleterious effects of the oxygen in the sintering atmosphere were also discussed in detail.

**Introduction**

The discovery of superconductivity at 26 K in the iron oxypnictide LaFeAs(O, F) by Hosono *et al*. [1] in 2008 soon afterwards lead to the development of iron-based superconductor families with different crystal structures, generally referred to as "1111" for REFeAs(O, F), "122" for AEFe$_2$As$_2$ [2] and AEFe$_2$Se$_2$ [3], "111" for LiFeAs [4] and "11" for (Fe(Se, Te)) [5]. Here RE denotes rare earth and AE denotes alkali earth. The REFeAs(O, F) superconductors, in which $T_c$ is over 50 K when La is replaced by Sm [6], Gd [7] or Tb [8], show very high upper critical fields $H_{c2}$ of about 300 T [9] and high intragrain critical current densities ($J_c$) over $10^6$ A cm$^{-2}$ (5 K, 0 T) [10]. However, the REFeAs(O, F) superconductors do share several properties with cuprates, such as high anisotropy, short coherence length [11], low carrier density [12], significant evidence for granularity and low intergranular $J_c$ [13,14]. The superconducting properties of the polycrystalline REFeAs(O, F) materials need to be greatly improved by optimal preparation conditions. Recently, many efforts were devoted to enhance the sample quality [14-16] through special techniques. The phase formation of LaFeAs(O, F) compound was studied in detail [17].

Since the REFeAs(O, F) superconductors contain elements that can be easily oxidized, ultra high vacuum and the presence of protective inert gases seem mandatory during preparation. The understanding of the detrimental effects of air on the formation and properties of REFeAs(O, F) systems is important in understanding the superconducting mechanisms and with regards to applications. However, no related investigations were yet reported. In this paper, we present detailed studies on a series of SmFeAsO$_{1-x}$F$_x$ samples sintered in quartz tubes filled with air of different pressures. The effects of the sintering atmosphere on the phase formation and superconductivity were investigated by means of crystal structure and resistivity measurements.

**Experimental**

The SmFeAsO$_{1-x}$F$_x$ samples were prepared by the two-step solid state reaction method. The precursor SmAs was first synthesized by reacting Sm and As chips in a vacuumed quartz tube at 500℃ for 10 h and then 750℃ for 10 h. The mixture of starting materials SmAs, Fe, Fe$_2$O$_3$ and FeF$_2$ powders with the nominal stoichiometric ratio of SmFeAsO$_{0.8}$F$_{0.2}$ were ground thoroughly and pressed into pellets all weighing about 1 g, then sintered in quartz tubes of the same volume at 1160℃ for 45 h. The alumina crucibles were used to hold the pellets to avoid contact with the quartz. The tubes were filled with atmosphere then vacuumed slowly to achieve required air pressure, $P_{air}$, which was monitored by a ZDF-5227-type vacuum gauge. To investigate the effects of the sintering atmosphere on the phase formation, the pellets made of SmAs and $FeF_2$ powders with the ratio of 10:1 were sealed in the quartz tubes filled with atmosphere, and the pellets made of SmAs, $Fe_2O_3$ and $FeF_2$ powders with the ratio of 6:3:2 were sealed in the quartz tubes with $P_{air}$=0.001 Pa. These pellets were sintered at 1100℃ for 20 h. The tube volumes were the same as those used to prepare the $SmFeAsO_{1-x}F_x$ samples.

Powder X-ray diffraction (XRD) was performed on a MAX-RC-type diffractometer with Cu-K$\alpha$ radiation from 2$\Theta$=20-80°. The calculations for the crystal parameters were performed on the basis of a least-square fit using the Checkcell programs. The temperature dependence of resistivity was measured by the standard four-probe method on a Quantum Design PPMS.

**Results and discussion**

The phase identification and crystal structure investigation were preformed on the samples sintered under various $P_{air}$. Figure 1 shows the XRD patterns normalized by the strongest peaks. The diffraction peaks can be well indexed on the basis of tetragonal ZrCuSiAs-type structure, confirming the main phases are $SmFeAsO_{1-x}F_x$. The impurities of SmOF and FeAs were detected in all the samples. The impurity SmAs, however, was only observed in samples sintered under vacuum (0.001 and 0.55 Pa). The impurity $Sm_2O_3$ was only observed in the sample sintered under 95 kPa, indicating SmAs was oxidized when $P_{air}$ was closed to the atmosphere. It is worth mentioning that there were brown deposits on the pellet surfaces and the quartz tubes when $P_{air}$<90 kPa, while the deposits were white acicular crystals as $P_{air}$=90 and 95 kPa. This phenomenon suggests that $FeF_2$ vaporized out of the samples during sintering, and $FeF_2$ seems not to react with the quartz heavily under vacuum. When $P_{air}$ reaches about 90 kPa, $FeF_2$ is oxidized by the oxygen ($O_2$) in the tubes. The reaction of $FeF_2$ is very likely to affect the fluorine (F) doping level, which can be checked by the crystal parameters [6]. From Figure 1, the crystal parameters $a$ and $c$ of the samples sintered under different $P_{air}$ were calculated as shown in Figure 2 in semilogarithmic scale. Both $a$ and $c$ increase with the increasing of $P_{air}$, indicating the decrease of F doping level. Moreover, $a$ and $c$ first increase slowly in a wide range of $P_{air}$, then increase dramatically when $P_{air}$ is over 75 kPa. Results suggest that F doping becomes more sensitive when $P_{air}$ reaches certain threshold. This can be verified by the $P_{air}$ dependence of $T_c$.

The transport properties of these samples were investigated. Figure 3 shows the temperature dependence of the resistivity, $\rho$, from 5 to 60 K. The transition temperature $T_c$ slowly decreases as $P_{air}$ increasing from 0.001 to 100 Pa, and dramatically shifts to lower temperature when $P_{air}$ reaches 75 kPa. This result is in agreement with the increase in $a$ and $c$ shown in Figure 2. The inset of Figure 3 shows $\rho$ versus temperature of the samples sintered under $P_{air}$=90 and 95 kPa. The 95 kPa $\rho(T)$ curve shows a spin-density-wave (SDW) instability at about 150 K as the parent compound [18]. The abnormality in the $\rho(T)$ curve of 90 kPa sample shifts to lower temperatures, indicating the SDW is suppressed by the F doping. According to the phase diagram [19], the actual F doping level in the 90 kPa sample is less than 0.037 to induce superconductivity. The mole fraction of $O_2$ in the quartz, $M_{O2}$, was calculated. $P_{air}$ was converted to $R$, where $R = \dfrac{M_{O2}}{M_{vac}} \times 100\%$. Here $M_{vac}$ is the mole number of oxygen vacancies in the samples (0.2 in $SmFeAsO_{0.8}F_{0.2}$). The residual resistivity ratio (RRR) was calculated as $\rho(300\ K)/\rho(52\ K)$. Figure 4 shows RRR versus both $P_{air}$ and $R$ in semilogarithmic scale. The RRR shows a shoulder with increasing $R$, and then decreases quickly when $R$ reaches 39%. The decrease of RRR together with the increase of normal-state resistivity indicates the increase of impurity scattering with increasing $R$.

Lastly, the effects of sintering atmosphere on the superconducting transition were studied. The onset $T_c$ was determined by 90% normal-state $\rho$. The transition width, $\Delta T_c$, was defined as $\Delta T_c = T(onset) - T(\rho = 0)$. $T_c$ and $\Delta T_c$ versus both $P_{air}$ and $R$ were also listed in Figure 4. The inset of Figure 4 shows the magnification in the range of $R$= 38% to 50%. The $R$ dependence of $T_c$ is in agreement with the RRR data, and the crystal parameter results in Figure 2. $T_c$ with the highest value of 51.48 K (0.001 Pa) drops to zero when $R$ is about 47%. The increase of $\Delta T_c$ from 2~3 K to 13.56 K indicates that the F doping becomes more inhomogeneous with increasing $R$. The reason may be that F reacts on the sample surface with $O_2$ in quartz tubes, producing an F-content gradient inside the sample.

It is assumed that the sintered samples were mainly affected by the air pressures and $O_2$ in the quartz tubes, because no nitride was observed in the XRD patterns, and the mole fraction of the other components, such as $CO_2$ and the moisture, were negligible compared to the pellets. The effects of air pressure are less pronounced since the samples sintered in vacuum underwent significant F vaporization but still have high $T_c$. The presence of $O_2$ may oxidize the starting materials, producing impurities of SmOF and $Sm_2O_3$, consuming F and reducing available oxygen vacancies [20]. The XRD patterns of the $SmAs+FeF_2$ pellets is shown in the bottom curve of Figure 5, with the peaks identified as SmAs and a small amount of SmOF. The reaction may be described as:

$$2SmAs + FeF_2 + O_2 => 2SmOF + FeAs_2 \tag{1}$$

Results here confirm that $O_2$ oxidizes the starting materials, consumes F and the other elements, and thus increases the relative oxygen ratio in the samples.

However, the $R$ values of the samples with $P_{air}$=90 and 95 kPa are only 47% and 50%, respectively. After reaction (1), there should still be about 50% of the residual vacancies and F to induce superconductivity. The fact that the sample of $P_{air}$=0.001 Pa also has impurity peaks suggests that impurities can form through the reactions between the starting materials without $O_2$. This was confirmed by the XRD pattern of the SmAs+$Fe_2O_3$+$FeF_2$ pellets sintered under 0.001 Pa in the upper curve of Figure 5, with the peaks identified as SmOF, FeAs and a small amount of $Fe_2As$. The reaction may be described as:

$$6SmAs + 3FeF_2 + 2Fe_2O_3 => 6SmOF + 5FeAs + Fe_2As \tag{2}$$

According to the results in Figure 5, it is postulated that when sintering under vacuum, F is lost due to the formation of SmOF and $FeF_2$ vaporization, the latter also causes an Fe deficiency, leading to the presence of extra SmAs, as observed in Figure 1. According to the phase diagram [21], the residual F doping level is about 0.08~0.1 to maintain a $T_c$ near 50 K. However, as the concentration of $O_2$ increases, $O_2$ reacts with SmAs and $FeF_2$ to form SmOF, leading to the increase of impurities supported by the RRR results, and the decreased sample homogeneity, which can be seen from $\Delta T_c$. The F was consumed to an insufficient doping level to induce superconductivity. Further investigations are needed to improve the sample quality by controlling the F vaporization and by reducing parasitic reactions to achieve a high level and homogeneous F doping.

**Conclusions**

In summary, the effects of sintering atmosphere on the SmFeAsO$_{1-x}$F$_x$ compound were systematically investigated. The evolution of the superconductivity with the concentration of oxygen was observed. Results indicate that fluorine is lost due to vaporization, reactions with starting materials and reactions with oxygen. A vacuum environment prevents oxidation through eliminating the presence of O$_2$. Techniques that can control the fluorine vaporization and techniques that can reduce the parasitic reactions are needed to improve the sample quality.

**Acknowledgments**

This work was supported by the scientific research foundation of graduate school of Southeast University (Grant No. YBJJ0933), and by the National Basic Research Program of China (973 Program, 2011CBA00105).

**Figure Captions**

Figure 1. X-ray powder diffraction patterns of the SmFeAsO$_{1-x}$F$_x$ samples sintered under different air pressures.

Figure 2. The lattice constant $a$ and $c$ of the samples sintered under different air pressures in semilogarithmic scale. The solid and dash lines are guides to the eyes.

Figure 3. The temperature dependence of the resistivity of the samples sintered under different air pressures from 5 to 60 K. The inset shows the temperature dependence of the resistivity of the samples sintered under 90 kPa and 95 kPa from 5 to 300 K.

Figure 4. RRR, $T_c$ and $\Delta T_c$ versus both air pressure and $R$ in semilogarithmic scale. Here $R$ is the ratio of the mole fraction of oxygen in quartz tubes versus the mole fraction of oxygen vacancies in the samples, and RRR=$\rho$(300 K)/$\rho$(52 K). The inset shows the magnification in the range of $R$= 38% to 50%. The solid and dash lines are guides to the eyes.

Figure 5. The XRD patterns of the pellets made of SmAs+Fe$_2$O$_3$+FeF$_2$ and SmAs+FeF$_2$ in upper and bottom, respectively.

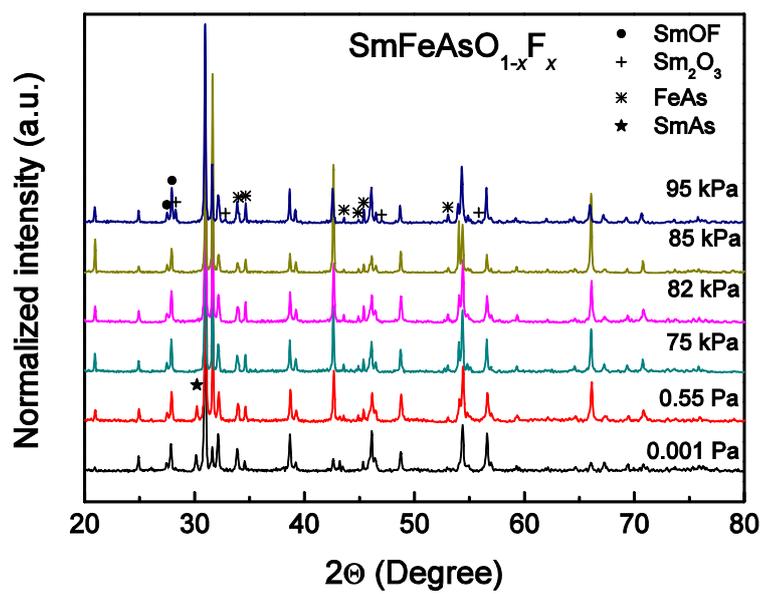

Figure 1 Y. Ding *et al.*

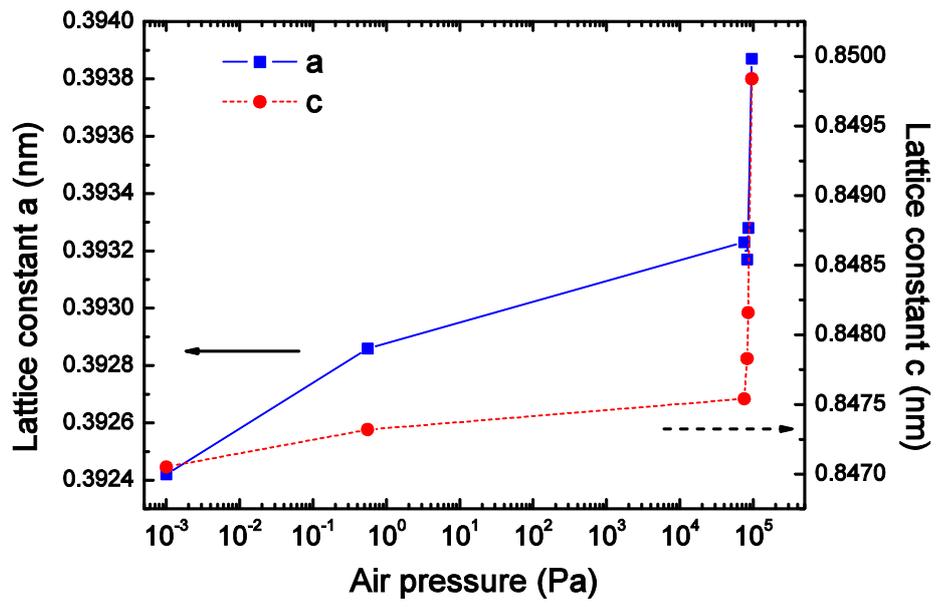

**Figure 2 Y. Ding** *et al.*

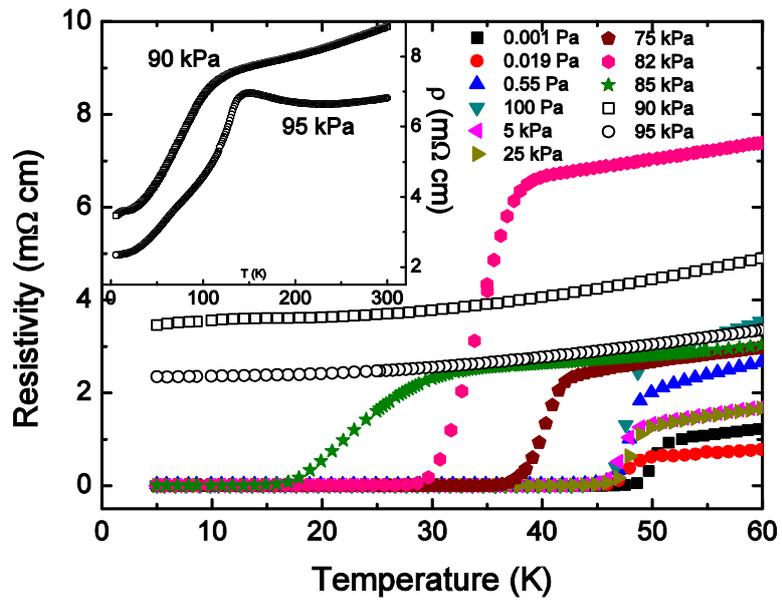

**Figure 3 Y. Ding *et al.***

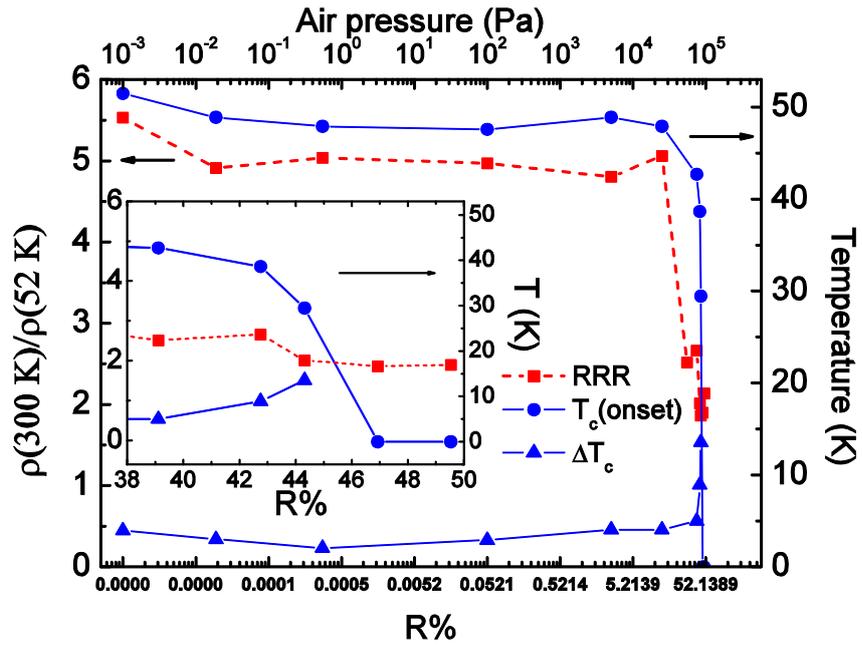

**Figure 4** Y. Ding *et al.*

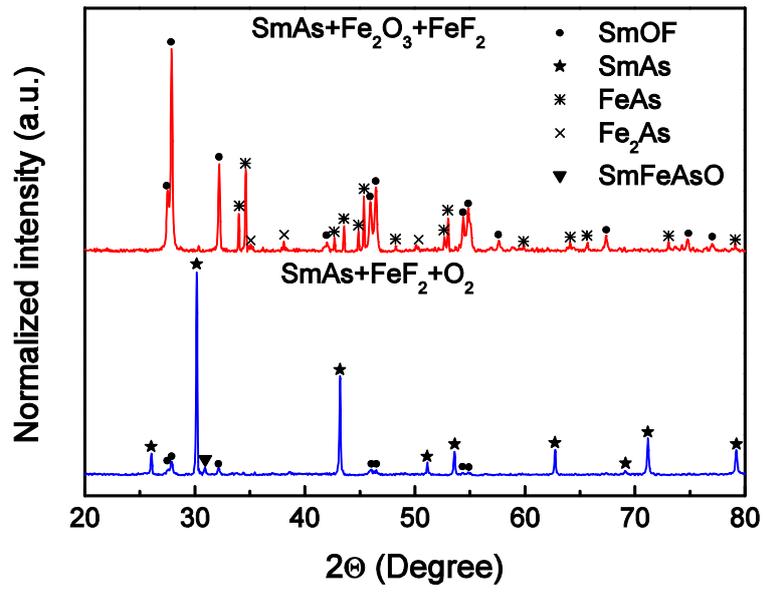

**Figure 5** Y. Ding *et al.*